\documentclass[aps,preprintnumbers,amsmath, amssymb]{revtex4}

\usepackage{float}
\usepackage{graphics,epsfig}
\usepackage{graphicx}
\usepackage{dcolumn}
\usepackage{bm}
\usepackage{braket}

\newcommand{\beq}{\begin{equation}}
\newcommand{\eeq}{\end{equation}}
\newcommand{\be}{\begin{equation}}
\newcommand{\ee}{\end{equation}}
\newcommand{\beqa}{\begin{eqnarray}}
\newcommand{\eeqa}{\end{eqnarray}}
\newcommand{\beqar}{\begin{eqnarray*}}
\newcommand{\eeqar}{\end{eqnarray*}}
\newcommand{\bea}{\begin{eqnarray}}
\newcommand{\eea}{\end{eqnarray}}

\def\ocal{{\mathcal{O}}}

\newcommand{\dd}{\textrm{d}}

\numberwithin{equation}{section}

\begin{document}
\baselineskip=0.8 cm

\title{{\bf Holographic flows with scalar self-interaction toward the Kasner universe  }}
\author{Yong-Qiang Wang$^{1,2,3}$, Yan Song$^{1,2}$, Qian Xiang$^{1,2}$, Shao-Wen Wei$^{1,2,3}$, Tao Zhu$^{4}$, Yu-Xiao Liu$^{1,2,3}$\footnote{liuyx@lzu.edu.cn, corresponding author}}
\affiliation{
$^1$Institute of Theoretical Physics, Lanzhou University, Lanzhou 730000, China\\
$^2$Research Center of Gravitation, Lanzhou University, Lanzhou 730000, China\\
$^3$Key Laboratory for Magnetism and Magnetic of the Ministry of Education, Lanzhou University, Lanzhou 730000, China\\
$^4$Institute for Theoretical Physics and Cosmology, Zhejiang University of Technology, Hangzhou, 310023, China}

\vspace*{0.2cm}
\begin{abstract}
\baselineskip=0.6 cm
\begin{center}
{\bf Abstract}
\end{center}
Considering a thermal state of the dual CFT  with  a uniform deformation
by a  scalar operator, we study a  holographic renormalization group
flow at nonzero temperature  in the bulk  described by the Einstein-scalar field theory with the self-interaction term
$\lambda \phi^4$ in asymptotic anti-de Sitter  spacetime. We show that the holographic flow with the self-interaction term  could run smoothly through the event horizon of a black hole and deform the Schwarzschild singularity to a  Kasner universe  at late  times.  Furthermore, we
also study the effect of the scalar self-interaction on the deformed near-singularity  Kasner exponents and  the relationship between entanglement velocity and Kasner singularity exponents at late times.

\end{abstract}

\maketitle
\newpage
\vspace*{0.2cm}

\section{Introduction}
According to general relativity, a gravitational singularity exists in the
interior of a black hole, which contains a huge mass in an infinitely small space. Due to that the infinity of matter density and gravity could lead to infinite spacetime curvature, all laws of physics as we know   break down at a singularity. Because of the existence of black hole singularity, the study of  the black hole interior is very challenging.
In recent years, the anti-de Sitter/conformal field theory (AdS/CFT) correspondence \cite{Maldacena:1997re,Witten:1998qj,Aharony:1999ti} has been used to investigate the black hole interior.
 It has been found \cite{Maldacena:2001kr,Kraus:2002iv,Fidkowski:2003nf,Kaplan:2004qe,Brecher:2004gn,Festuccia:2005pi} that  by studying the analytic behaviors of the correlation functions of conformal field theory,  one could obtain the information about the singularity extracted from the corresponding correlation functions.
 For a dynamical spacetime geometry, the entanglement entropy is also a possible way  to probe the interior of a black hole \cite{Hubeny:2002dg,AbajoArrastia:2010yt,Hubeny:2012ry,Hartman:2013qma,Liu:2013iza,Hubeny:2013dea}. Especially,
 at late times the linear growth of the entanglement entropy could be directly related to  the growth of the black hole interior measured along with
a critical  spatial slice \cite{Hartman:2013qma}.

Recently,  the model of a massive, real scalar field minimally  coupled  to the four-dimensional Einstein gravity with a negative cosmological constant was investigated in \cite{Frenkel:2020ysx}. Considering  that  the thermal state of the dual CFT field could be
deformed by a relevant scalar operator sourced  by the scalar field on the boundary,  the authors solved numerically the equations of motion  from an
AdS boundary as the UV, to the timelike  singularity inside the black hole as the IR, and found that  the
 generic behavior of this family of numerical solutions near the black
hole singularity  could be described by a one parameter family of homogeneous, anisotropic
so-called Kasner spacetime. Moreover, with the vanishing of deformation, the IR geometry has a timelike cousin of the Schwarzschild singularity and  is the less generic case of the Kasner universe. Soon afterward the study of holographic flows from CFT to Kasner universe was extended to the case of charged black holes \cite{Hartnoll:2020rwq}.
It is interesting to see that a relevant deformation of the dual
CFT with a neutral scalar operator  would result in the formation of a charged black hole without a Cauchy horizon. Furthermore, in \cite{Hartnoll:2020fhc} holographic flow inside the horizon of  holographic superconductors  was studied, and complex dynamical behaviors in the interior of the holographic superconductors were found.

Until now, the studies of holographic flow from the  UV to the IR only focus on the case of the free scalar field.  Moreover, in \cite{Hartnoll:2020fhc} the authors thought that in some epochs  of the dynamics,
 holographic flows inside the horizon of  holographic superconductors    are likely sensitive to the scalar potential, and it is worth studying more general scalar potentials.
In this paper, we  investigate  a  holographic renormalization group
flow at nonzero temperature in the bulk  described by the Einstein-scalar field theory with the self-interaction term
$\lambda \phi^4$ in asymptotic  AdS spacetime,
and study the effect of  scalar self-interaction on the  near-singularity  Kasner exponents.
Furthermore, as a method to probe the black hole interior,   entanglement velocity for the black hole with the Kasner singularity interior at late times is also investigated.

 The paper is organized as follows. In Sec. \ref{sec2}, we introduce the model of the Einstein-scalar field theory with the self-interaction term
$\lambda \phi^4$ in asymptotic AdS spacetime and explore the ansatz of metric and matter field. We also analyze the boundary conditions. In Sec. \ref{sec3},  we introduce the numerical   method and show the numerical results for a class of  holographic flows from the AdS boundary to the Kasner singularity. Moreover,  the effect of the scalar self-interaction on the deformed near-singularity  Kasner exponents and   entanglement velocity for the black hole at late times are studied.  The conclusion and discussion are given in the
  last section.
\section{Set up}\label{sec2}
We consider the action of a massive, real scalar field minimally coupled  to the $3 + 1$ dimensional Einstein gravity with a negative cosmological constant,
  which is written as
\begin{equation}
\label{eq:action}
S = \int\!\dd^4x\,\sqrt{-g} \, \left [ \frac{1}{2 \kappa_4^2} \left( R -\Lambda\right) - \frac{1}{2} \left(g^{\mu\nu} \nabla_\mu \phi \nabla_\nu \phi + m^2 \phi^2+\lambda \phi^4 \right) \right]\,,
\end{equation}
where $\kappa_4$ is the four-dimensional gravitational constant and $\Lambda = - \frac{6}{L^2}$ is the cosmological constant with $L$ being the AdS radius.
$R$ and $\phi$ are the  Ricci scalar and real scalar
field respectively, and the latter has  a  quartic self-interaction.
The constants $m$ and $\lambda$ represent the  mass of the scalar field and  the interaction parameter, respectively.
The  parameter $\lambda$  controls the
strength of this self-interaction, and
the self-interacting scalar field  reduces to the free one when $\lambda=0$.
The Einstein and scalar field  equations can be derived from the above action (\ref{eq:action}) as
\begin{subequations}
\begin{align}
R_{\mu\nu}+\frac{3}{L^{2}}g_{\mu\nu}&=\frac{\kappa_4^2}{2 }\left(2\nabla_{\mu}\phi\nabla_{\nu}\phi +g_{\mu\nu} (m^{2}\phi^2+\lambda \phi^4 )\right)\,, \label{eq:einsteins}
\\
\nabla_\mu\nabla^\mu \phi&=m^{2}\phi+2 \, \lambda \,\phi^3 \,.
\label{eq:scalar}
\end{align}
\label{eqs:motion}
\end{subequations}
For simplicity, we will set the AdS radius to one  and  $ \kappa_4^2= 1/2$. Note that in the four-dimensional spacetime the values of mass need to satisfy the Breitenlohner-Freedman (BF)
bound of $m^2 \geq -9/4$ \cite{Breitenlohner:1982bm}, and we will set $m^2 = -2$.

In order to obtain the plane-symmetric hairy black hole solutions, following the  conventions in \cite{Frenkel:2020ysx} we
choose the metric ansatz  in the form of
\be\label{eq:adsBH}
ds^2 = \frac{1}{r^2} \left( - f(r) e^{-\chi(r)} dt^2 + \frac{dr^2}{f(r)} + dx^2 + dy^2 \right) \,,
\ee
together with the scalar field
\be\label{eq:adsBH1}
\phi=\phi(r)\,,
\ee
where the  radial coordinate $r \in (0, \infty)$.  The  asymmetrical AdS boundary  and the singularity inside the event horizon
are fixed at $r \to 0$ and $r \to \infty$, respectively.  The horizon radius  $r_+$ is defined through the requirement that $f(r_+) = 0$.
 When the scalar field vanishes, there exists a solution of the  planar Schwarzschild-AdS black hole with   $\chi = 0$  and $f = 1 - (r/r_+)^3$.

 As shown in \cite{Kasner, Belinski:1973zz}, inside the horizon,
the near-singularity behavior of  the metric (\ref{eq:adsBH}) has the form of the Kasner universe,
\be\label{eq:kasner}
ds^2 \sim - d\tau^2 + \tau^{2 p_t} dt^2 + \tau^{2 p_x} \left(dx^2 + dy^2 \right) \,,
\ee
with the near-singularity behavior of the scalar field given by
 \be\label{eq:kasner1}
\qquad \phi(r) \sim - \sqrt{2} p_\phi \log \tau \,.
\ee
Here $\tau$  is a time function with  the singularity at $\tau= 0$. Setting $d\tau=(-r^2 f)^{-1/2}dr$, the metric of the plane-symmetric  black hole (\ref{eq:adsBH}) can take the form (\ref{eq:kasner}).
 The parameters $p_\phi,  p_t, p_x$ are  known as
the Kasner exponents, which are functions of the spatial coordinates $t,x,y$ and must satisfy the Kasner relations $p_t + 2 p_x = 1$ and $p_\phi^2 + p_t^2 + 2 p_x^2 = 1$.
For the Schwarzschild-AdS black hole,  the Kasner-type singularity has the exponents $p_t = - \frac{1}{3}, p_x = \frac{2}{3}$   and $p_\phi = 0$.
 The solution (\ref{eq:adsBH}) is asymptotically AdS, which describes an RG flow from the
AdS boundary as the UV, to the timelike Kasner singularity as the IR.

In order to obtain the numerical solution from the AdS boundary to the near-singularity inside a black hole horizon, we will choose the  ingoing coordinates as follows
\be\label{eq:ansatz}
ds^2 = \frac{1}{r^2} \left( - f(r) e^{-\chi(r)} du^2 + 2 e^{-\chi(r)/2} du dr + dx^2 + dy^2  \right) \,.
\ee
One advantage of this metric is that  the apparent singularity at the event horizon where $f(r_+) = 0$ is only a coordinate singularity which is not   physical.
Substituting the ansatz of the metric (\ref{eq:ansatz}) and the matter field (\ref{eq:adsBH1}) into the Einstein  and  scalar field equations (\ref{eqs:motion}), one can derive the equations
of motion as follows
\begin{subequations}
\begin{align}
\phi'' + \left(\frac{f'}{f} - \frac{2}{r} - \frac{\chi'}{2} \right)\phi' + \frac{2}{r^2 f} \phi   -\frac{2 \lambda \phi^3}{r^2 f} = &\, 0 \,, \\
\chi' - \frac{2 f'}{f} - \frac{\phi^2}{r f} - \frac{6}{r f} + \frac{6}{r} +\frac{\lambda \phi^4}{2 r f} = & \,0 \,, \\
\chi' - \frac{r}{2} (\phi')^2 = & \,0 \,.
\end{align}
\label{eqs:motionss}
\end{subequations}
To obtain the numerical solutions of the above equations  by integrating  from the AdS boundary to the near-singularity region, we require that $f=0$
at the event horizon $r = r_+$. Thus,
the Hawking temperature of the black hole is
\be\label{eq:temp}
T = \frac{|f'| e^{- \chi/2}}{4 \pi}|_{r=r_+} \,.
\ee
In addition, by analyzing
the equations of motion (\ref{eqs:motionss}) at the asymptotic boundary  ($r \to 0$) order by order, we
obtain the asymptotic expansions as follows
\begin{align}
\phi = \phi_o  r + \langle \ocal \rangle r^2 + \cdots \,, \; \quad
\chi  = \frac{\phi_o ^2}{4} r^2 + \cdots \,, \; \quad  f =  1 + \cdots \,. \label{eq:nearb}
\end{align}
According to AdS/CFT dictionary,  the first-order expansion
coefficient $\phi_0$ is the source in the dual field theory on the AdS
boundary, meanwhile,  the second-order one $\langle \ocal \rangle$
is   the corresponding expectation value of the operator $\ocal$.
Moreover,  solving
the equations of motion (\ref{eqs:motionss})  near the singularity $( r \to \infty )$, we
obtain the asymptotic behaviors in the  form  of
\be\label{eq:cs}
\phi = 2 c \, \log r + \cdots \,, \qquad
\chi = 2 c^2 \log r  + \cdots \,, \qquad
f = - f_1 r^{3 + c^2} + \cdots \,.
\ee
Here both $c$ and $f_1$ are  constants, and $c=0$ corresponds to the case of  the Schwarzschild-AdS black hole. Furthermore,
by setting the  coordinate $r^{(3 + c^2)} = 1/\tau^2$, we can write the spacetime near the singularity in form of the Kasner university metric form (\ref{eq:kasner}) with the Kasner exponents:
\be\label{eq:ppp}
p_x  = \frac{2}{3 + c^2} \,, \quad p_t = \frac{c^2 - 1}{3 + c^2} \,, \quad p_\phi = \frac{2 \sqrt{2} c}{3+ c^2} \,.
\ee

\section{Numerical results}\label{sec3}
In order to obtain the numerical solutions of  the  coupled equations (\ref{eqs:motionss}),
 the  integration over the whole region $0 \leq r \leq \infty$  should be divided into the following two steps. In the first step, we use the pseudospectral method based on the Chebyshev polynomials in the integration region $0 \leq r \leq r_+$. Our iterative process is performed by using the Newton-Raphson method, and the
relative error for the numerical solutions in this  process is estimated to be below $10^{-5}$. For details of the pseudospectral method, see Ref. \cite{Dias:2015nua}.  Of course, we can also solve  these  equations of motion by using the numerical shooting method
described in \cite{Hartnoll:2008vx,Hartnoll:2008kx}.
 In the second step, we consider the backward differentiation formula (BDF) for solving Eqs.  (\ref{eqs:motionss}) in the integration region $r_+ \leq r \leq \infty$, which  is a family of implicit methods for the numerical integration of ordinary differential equations.  The initial value for the second step is extracted from the  data on the event horizon derived from the first step.

\subsection{Holographic flows from UV to IR}
\begin{figure}[ht!]
    \centering
    \includegraphics[width=0.6\textwidth]{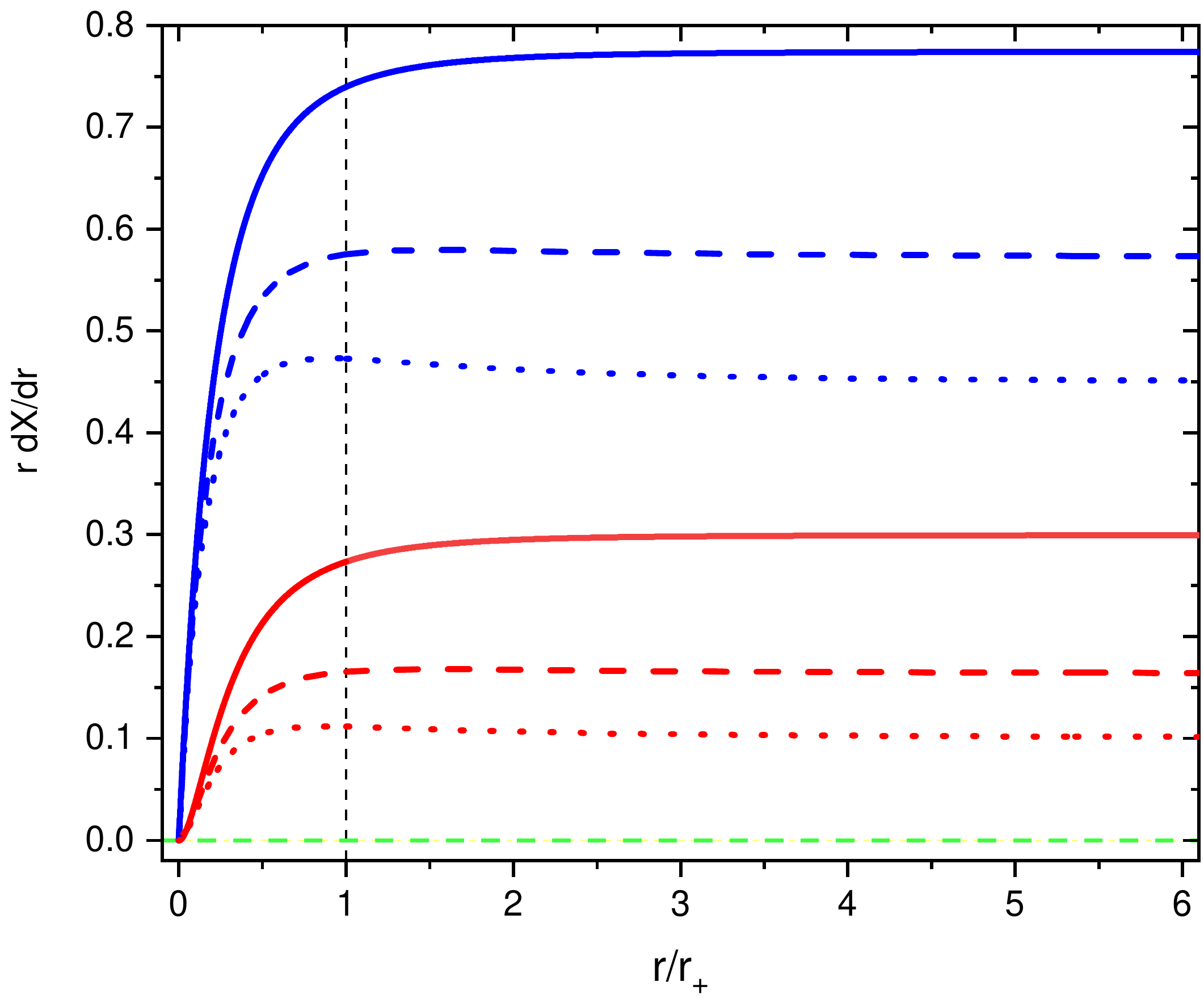}
    \caption{The holographic flow from the AdS boundary to the Kasner universe inside the event horizon.
    The blue  and red lines
correspond  to the functions $r \frac{d\phi}{dr}$ and  $r \frac{d\chi}{dr}$, respectively.  The solid, dashed, and dotted curves denote different self-interaction
parameters $\lambda=0, \lambda=0.1$ and $\lambda=0.2$, respectively.
    All of curves are fixed with the deformed parameter ${\phi_0}/{T}=13$.}
    \label{fig:kasner}
    \vskip -0.1in
\end{figure}

After obtaining  the numerical solutions of the deformed black hole with scalar hair,  we
 study how  the scalar self-interaction parameter $\lambda$ affects  the deformed near-singularity  Kasner exponents.
As an example, we show the typical results of our numerical program
for  $r \frac{d\phi}{dr}$ and  $r \frac{d\chi}{dr}$ as two functions of $r$ for a fixed  dimensionless parameter $\phi_0/T=13$ in Fig. \ref{fig:kasner}.  The blue  and red lines
correspond  to the functions $r \frac{d\phi}{dr}$ and  $r \frac{d\chi}{dr}$, respectively.  The dashed  and dotted curves denote different self-interaction
parameters $\lambda=0.1$ and $\lambda=0.2$, respectively. When taking the parameter $\lambda=0$ (solid curves), we  reproduce holographic flow from AdS to a Kasner cosmology, which was first obtained in \cite{Frenkel:2020ysx}. The horizontal dashed green line along the $r$ coordinator corresponds  to the planar Schwarzschild-AdS black hole. The vertical dashed black line
indicates the location of the event horizon.   In the  large $r$   limit (near singularity),  the values of $r \frac{d\phi}{dr}$ and  $r \frac{d\chi}{dr}$  are  close  to  different non-zero constants,  which is the universal feature  with the different self-interacting parameters $\lambda$. Moreover, we could see that the
 asymptotic values of the holographic flows from CFT to the Kasner universe decrease  with
 the scalar self-interaction $\lambda$.

\begin{figure}[ht!]
    \centering
    \includegraphics[width=0.6\textwidth]{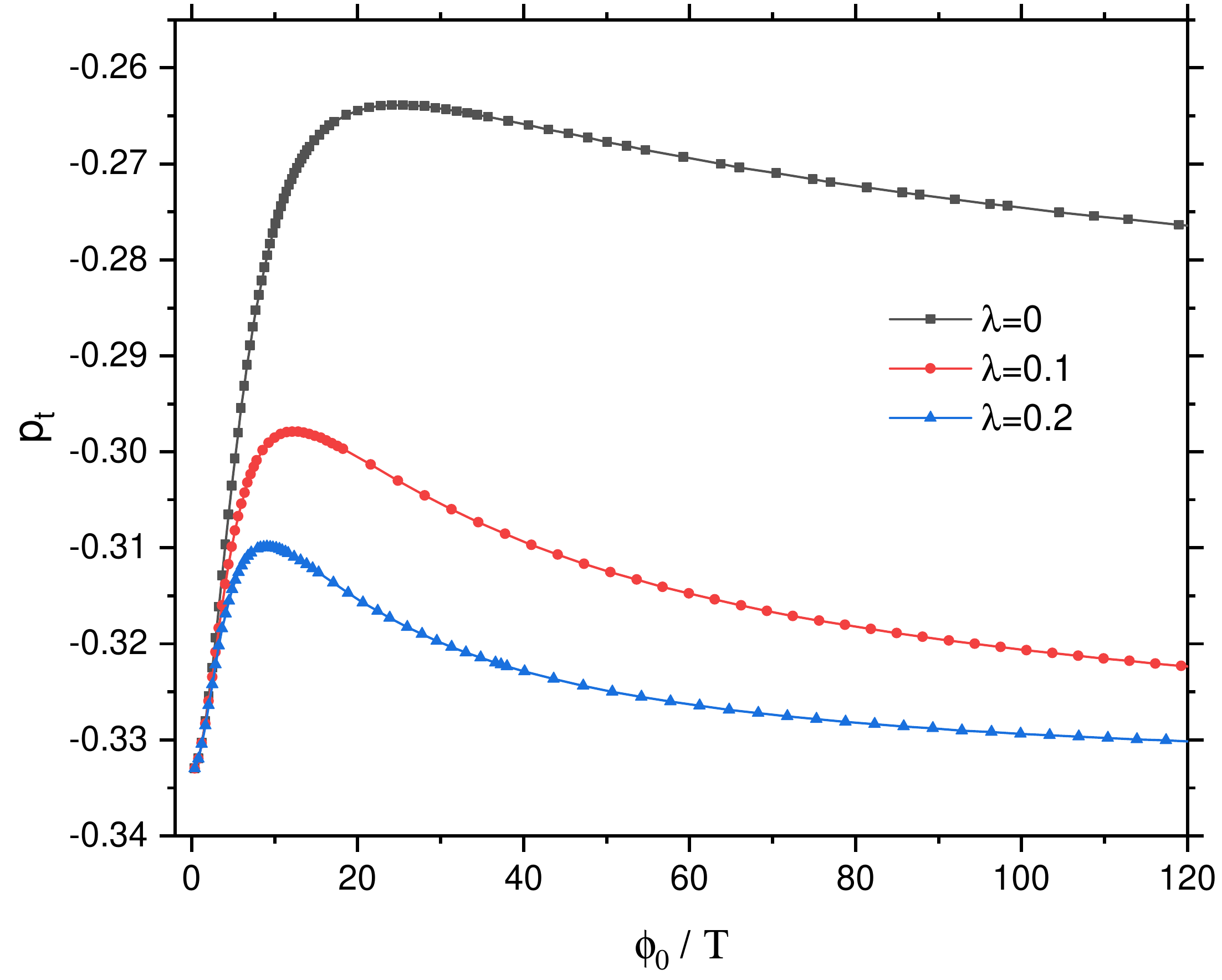}
    \caption{The Kasner exponent $p_t$  as a function of
the deformed parameter ${\phi_0}/{T}$ with different scalar self-interaction parameters $\lambda$. From top to bottom, the black, red  and blue curves denote the self-interaction
parameters $\lambda=0, \lambda=0.1$ and $\lambda=0.2$, respectively.}
    \label{fig:kasner1}
    \vskip -0.1in
\end{figure}

From the   asymptotic behavior of the solutions of Eq. (\ref{eq:cs}), we observe that  the
 asymptotic values of the holographic flows could be related to the Kasner exponents (\ref{eq:ppp}).
 In Fig. \ref{fig:kasner1}, we  show how the Kasner exponent $p_t$ varies as a function of
the deformed parameter ${\phi_0}/{T}$ with different scalar self-interaction parameters $\lambda$.
From top to bottom,  we plot three curves with  $\lambda=0$ (black color
), $\lambda=0.1$ (red color), and $\lambda=0.2$ (blue color), respectively.
 At ${\phi_0}/{T}=0$, all of the Kasner exponents in these three curves  reduce to the usual Schwarzschild-AdS singularity with the value of $p_t= -1/3$.
With the increase of the  deformed parameter  ${\phi_0}/{T}$, the Kasner exponent  $p_t$ increases firstly
and then  reaches a maximum point. Further increasing the value of ${\phi_0}/{T}$, the Kasner exponent begins to decrease.
When the curves go  into the  range of large deformed parameter ${\phi_0}/{T}$, the numerical error begins to increase and a finer mesh is required. However,
it is certainly possible that the  curve
will eventually approach the Schwarzschild singularity value of $p_t= -1/3$ at a much larger  deformed parameter ${\phi_0}/{T}$. Moreover, we could see that
at a fixed deformed parameter ${\phi_0}/{T}$,  the Kasner universe   exponent decreases in the conditions of higher scalar self-interactions  $\lambda$.
As $\lambda\rightarrow\infty$,  it can be predicted that the Kasner exponent $p_t$ would return to the Schwarzschild singularity with the value of -1/3.
It is noteworthy that the  Kasner exponents with the different self-interaction parameters $\lambda$ are close in the  range of small deformation ${\phi_0}/{T}$, the reason is that the self-interacting term $\lambda\phi^4$
is a small quantity in the range of small deformation due to its high order and thus  has little effect on the deformed near-singularity Kasner exponents.
 While,  in the  range of large deformation ${\phi_0}/{T}$,  the self-interacting term is big enough to affect  the Kasner exponents.
	
\subsection{ Probes of the Kasner exponent  by  entanglement entropy }
In this subsection, we would use the entanglement entropy as a probe  of the black hole interior to study the   Kasner singularity.
It is well known that the entanglement entropy of the dual field could be  obtained in holographic models
as the area of the minimum  surface  extended into the bulk  with the same AdS boundary of the quantum system  \cite{Ryu:2006bv,Hubeny:2007xt,Lewkowycz:2013nqa}.    Hartman and Maldacena in \cite{Hartman:2013qma} showed that
 at late times of the dynamical system, the extremal surface eventually stops expanding on a specific critical  slice  inside the horizon and can not get closer to the
singularity, and the  entanglement entropy increases linearly with time.
The linear growth of the entanglement entropy could be directly related to  the growth of the black hole interior measured along
this critical surface. This growth could define a so-called entanglement velocity, which is sensitive to the
black hole interior, but not to the near-singularity region. In our model,  at late time   the entanglement velocity $v$ is written as
\be\label{ev}
 v^2 = r_+^4 \left. \frac{|f| e^{-\chi} }{r^4}\right|_{r=r_\text{crit}}. \,
\ee
Here  $r=r_\text{crit} $ is the  radius of the critical surface, in which  $-f e^{-\chi}/r^4$ has a maximum inside the horizon. The formula of the entanglement velocity in Eq. (\ref{ev}) is the same as the case of the free scalar field in \cite{Frenkel:2020ysx}.

\begin{figure}[ht!]
    \centering
    \includegraphics[width=0.6\textwidth]{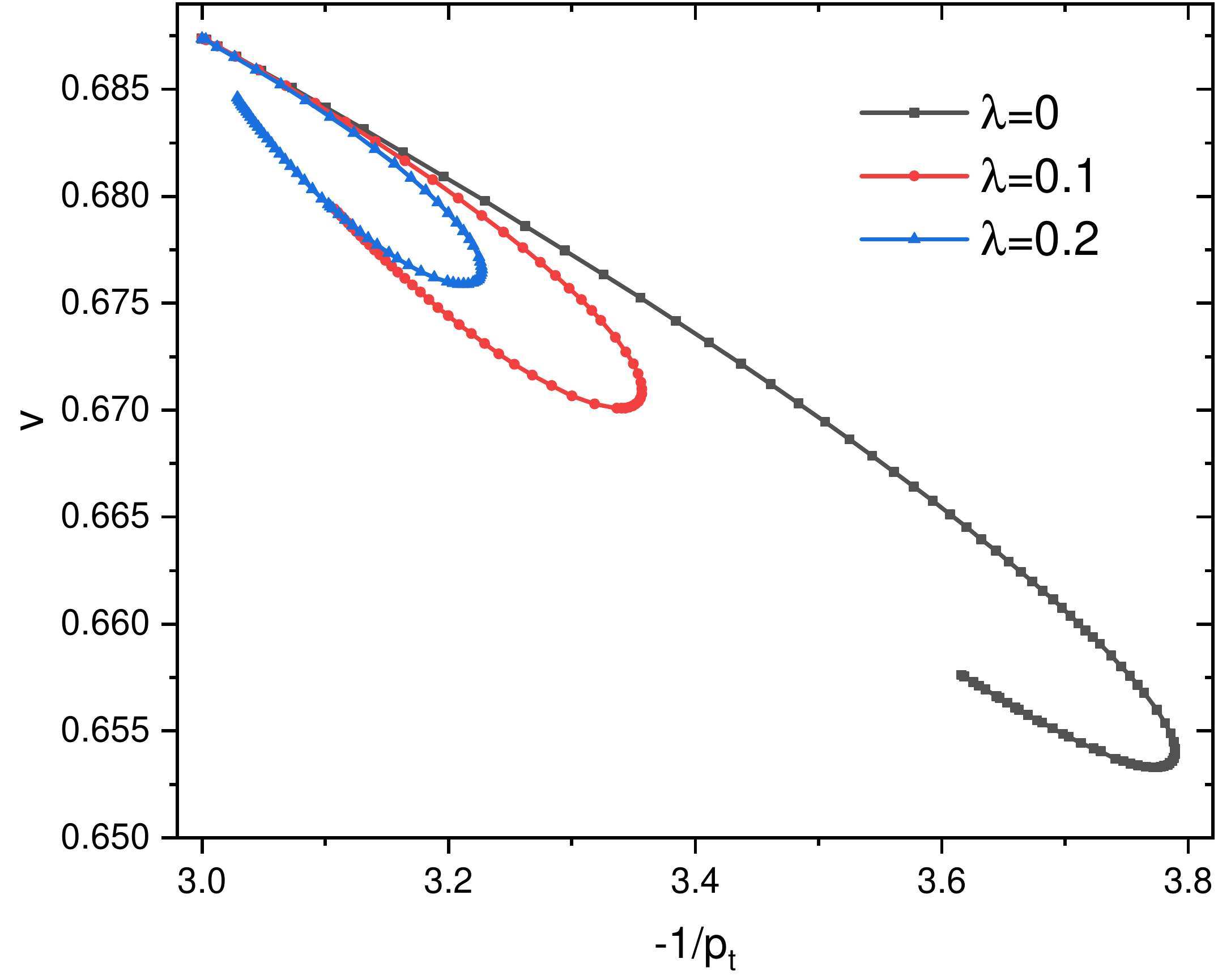}
    \caption{The dependence of the entanglement velocity $v$  on $-1/p_t$. The black, red  and blue curves correspond to the self-interaction
parameters $\lambda=0, \lambda=0.1$ and $\lambda=0.2$, respectively.}
    \label{fig:kasner2}
    \vskip -0.1in
\end{figure}
In Fig. \ref{fig:kasner2},	we exhibit the entanglement velocity  $v$  as a function of $-1/p_t$ with the scalar self-interaction parameter $\lambda=0, \;0.1, \; 0.2$, denoted by the black,  red,  and blue
dotted lines, respectively.
 At $-1/p_t=3$, the values of the entanglement velocity in these three curves  reduce to the usual Schwarzschild-AdS singularity with the entanglement velocity $v=\sqrt{3}/2^{4/3}$.
With the increase of $-1/p_t$, the entanglement velocity decreases at first
and then it reaches a minimum value.
Afterwards, by re-decreasing $-1/p_t$, the value of the entanglement velocity slowly goes upwards and reach the Schwarzschild-AdS singularity in the end.
 Moreover, we could see that the
minimum value of the  entanglement velocity
increases when enlarging the scalar self-interaction $\lambda$.

\section{Conclusion and discussion}
In summary, we have studied a  holographic renormalization group
flow at nonzero temperature in the bulk  from the AdS boundary to  the Kasner universe. By numerically solving the gravitational dynamics in the ingoing coordinates,
we showed  the holographic flow with the self-interaction term  could run smoothly through the event horizon of the black hole and deform the Schwarzschild singularity to the
Kasner universe, where the Kasner exponent decreases  with the scalar self-interaction $\lambda$. Moreover, in the range of small  deformation ${\phi_0}/{T}$,  the scalar self-interaction term has  a smaller  effect on  the deformed near-singularity  Kasner exponents, and in the  range of large deformation,  the self-interacting term is big enough to affect  the Kasner exponents.
 Furthermore, we  used the entanglement entropy as a probe  of the black hole interior to study the  Kasner singularity and showed that the minimum  of  the entanglement velocity increases  with
 the scalar self-interaction $\lambda$.

 There are several interesting extensions of our work. The first one is   to investigate how
 a relevant deformation of the dual
CFT with a  self-interacting scalar operator  would result in the formation  of the Kasner singularity in a charged black hole interior,  and   study whether the charged black hole with the scalar self-interaction has  no Cauchy horizon similar as the case in \cite{Hartnoll:2020rwq}. The second one is to extend the study on  holographic flow inside the horizon of the holographic superconductors \cite{Hartnoll:2020fhc}  to the model with more general scalar potentials, and investigate how the  dynamical behavior in the  interior  of the holographic superconductors performs.
Finally, we are planning to study the holographic flow toward the black hole interior in the
model of the excited holographic superconductor \cite{Wang:2019caf,Wang:2019vaq},  and investigate the configuration  of  the  black hole with Kasner singularity in future works.

\textbf{Note added:} When we are submitting this paper to arXiv, we notice that there appears a paper on arXiv
\cite{Cai2009.05520}, in which the authors established a no inner-horizon theorem for black holes with charged scalar hair including the self-interaction term, and proved that in a general gravitational theory with a charged scalar field  both spherical and planar black holes with scalar hair have no inner Cauchy horizon.

\begin{acknowledgments}
Y.-Q. Wang would like to thank Jie Yang and Li Zhao   for  helpful discussions.  Parts of computations were supported by Supercomputing Center of  Lanzhou university.
This work was supported by the National Natural Science Foundation of  China under Grants. No. 11675064 and No. 11875175 and  the Fundamental Research Funds for the
Central Universities under Grant No. lzujbky-2019-ct06. T. Zhu was supported in part by the National Natural Science Foundation of China under Grant No.11675143, the Zhejiang Provincial Natural Science Foundation of China under Grant No. LY20A050002, and the Fundamental Research Funds for the Provincial Universities of Zhejiang in China under Grant No. RF-A2019015.

\end{acknowledgments}

\end{document}